\documentclass[aps,twocolumn,nofootinbib,preprintnumbers,citeautoscript,superscriptaddress]{revtex4}
\usepackage{amsmath,graphicx, amssymb, amsthm, times,float,xcolor}
\newcommand{\comment}[1]{}
\begin{document}
\title{Improvements to the Froissart bound from AdS/CFT}
\smallskip\smallskip 
\author{Ver\'{o}nica Errasti D\'{i}ez}
\email{vediez@physics.mcgill.ca}
\affiliation{Centre for High Energy Physics, Indian Institute of
Science, 
Bangalore 560012, India}
\affiliation{Physics Department, McGill University, 3600 University St, Montr\'{e}al, QC H3A 2T8, Canada}
\author{Rohini M. Godbole}
\email{rohini@cts.iisc.ernet.in}
\affiliation{Centre for High Energy Physics, Indian Institute of
Science, 
Bangalore 560012, India}
\author{Aninda Sinha}
\email{asinha@cts.iisc.ernet.in} 
\affiliation{Centre for High Energy Physics, Indian Institute of
Science, 
Bangalore 560012, India}

  \begin{abstract}
In this paper we consider  the issue of the Froissart bound on the high energy behaviour of total cross sections.
 This bound, originally derived using principles of analyticity of scattering amplitudes, is seen 
 to be satisfied by all the available experimental data on total hadronic cross sections.
 At strong coupling,  gauge/gravity duality
 has been used to provide some insights into this behaviour.
In this work, we find the subleading terms to the so-derived Froissart bound from AdS/CFT. We find that a
$(\ln \frac{s}{s_0})$ term
is obtained, with a negative coefficient. We see that the fits to the  currently available data confirm
improvement in the  fits due to the inclusion of such a term, with the appropriate sign.
\end{abstract}

\maketitle

\newpage

\section{Introduction}\label{intro}

The high energy behaviour of the total cross sections for the scattering of any two high energy particles has
been a
subject of great theoretical interest over many decades, beginning from Heisenberg~\cite{Heisenberg}.
The various bounds one obtains on the rate of the rise of these cross sections  are based just on
the analyticity properties of scattering amplitudes and do not use any knowledge of the underlying dynamics
of the interaction responsible for the scattering. The most important of all these has been the Froissart
bound~\cite{Froissart,Martin},
which states that at high energies, the total cross section for a scattering process $1+2\rightarrow f$
($f=\,\,$any final state) has an  upper bound $\sim(\ln^2\frac{s}{s_0})$. Here $\sqrt{s}$ is the centre
of mass energy and $\sqrt{s_0}$ is an energy scale.
Since the bound can be derived from very general physical  arguments, such as the  unitarity of
the $S$ matrix and certain analytical properties of the scattering amplitude, it has to be true
in any quantum field theory.
However, to derive this in the context of theories of strong interactions, like QCD, one
requires to handle the theory in the non-perturbative regime. As a result, there exist only models
and these  usually try to incorporate known properties of QCD, the modelling aspect involving assumptions and
ansaetze about the non-perturbative
regime~\cite{Achilli:2011sw,Grau:2009qx,Achilli:2007pn,
Ferreiro:2002kv,Block:1998hu,Carvalho:2007cf,Luna:2005nz,Fagundes:2015vba}.
In fact, analyses of~\cite{Grau:2009qx,Achilli:2009fk}
in the framework of the Bloch-Nordsiek improved eikonalised mini jet model~\cite{Godbole:2004kx} indicate a
direct
relationship 
between the Froissart bound at high energy and the dynamics of ultra soft gluons, i.e., the behaviour of
QCD in the far infrared.  An analysis of the high energy
behaviour of the available data on total hadronic cross sections, while being completely consistent with
the bound, seems to need in the fits the presence of a subleading $(\ln \frac{s}{s_0})$ term to the Froissart
bound. Various QCD (inspired) models use such a subleading term, though there is no theory with a proper
explanation of this subleading behaviour~\cite{Achilli:2007pn,Cudell:2001pn,Igi:2005jm,Block:2005ka}.

In this article, we attempt to seek a theoretical understanding of this subleading behaviour
shown by the data on total  cross sections at high energies, using the AdS/CFT
correspondence~\cite{Maldacena:1997re}.
It is known that the leading term (Froissart bound) can be derived using
AdS/CFT~\cite{Giddings:2002cd, Kang:2004jd, Kang:2005bj}. 
The model based on~\cite{Polchinski:2001tt} describes high energy scattering in a large-$N$,large-$\lambda$ 
gauge theory with broken conformal symmetry which is achieved by putting a cut-off in the infra-red (IR).
Here $N$ is the number of colours and $\lambda$ is the 't Hooft coupling.
When a point mass $m$ is placed on the IR boundary,
the perturbations in the AdS space can be big enough to create a black hole. The geometrical cross section
of such a black hole, whose radius is equal to the AdS radius, is the gravity dual of the maximum possible
scattering cross section in the field theory. In order to have a more realistic model,
which takes into account finite coupling corrections (in $1/N$ and $1/\lambda$), we can incorporate higher
curvature corrections in the dual gravity description. If we are working perturbatively in the couplings,
then the leading 
correction  of this type will be at four derivative order~\cite{Buchel:2008vz}. Such terms can be redefined
into a single Gauss-Bonnet term whose coefficient describes a $1/N$ correction -- see eg.~\cite{Buchel:2009sk}
for the dictionary relating the field theory and gravity in such a theory. As we will argue below, this result
will in fact hold for any higher curvature gravity dual based on the results in~\cite{ss}.
In~\cite{Kang:2004jd,Kang:2005bj} it is argued that the corrections due to string modes are exponentially
small, so here we will ignore them from the onset. However it is important to take 
into account $1/N$ and $1/\lambda$ corrections to ensure that the behaviour of the subleading terms
arising from Einstein gravity as additional contributions to the Froissart bound do not get significantly
modified.

In this paper, we will consider subleading corrections to the Froissart bound that arise from 
such an analysis using  the results of~\cite{Giddings:2002cd}. The incorporation of the
higher curvature terms in the dual gravity will enable us to see what the finite coupling effects are.
We show that the subleading term in that cross section is indeed $(\ln \frac{s}{s_0})$, as 
seems to be required by the fits. 
Our AdS/CFT analysis predicts that the coefficient of this subleading term should be negative.
Further, if we consider the  subleading term coming in the dual impact parameter from Einstein gravity,
in the dual field
theory there is an additional  correction to
the Froissart bound of the form $(\ln\frac{s}{s_0})(\ln\ln\frac{s}{s_0})$. A recent work~\cite{Nastase:2015ixa}
also found the same subleading term as a correction to the maximal impact parameter. At high
energies, we expect that the general structure of the subleading terms that we find from AdS/CFT
will be useful in getting better fits to experimental data. This is precisely what we find.

Finally, we use all available data (including LHC and cosmic ray data)
to conclude that the $(\ln \frac{s}{s_0})$ term  significantly improves the fit (the $F$ test is
used). Also, the data always chooses the coefficient of the $(\ln \frac{s}{s_0})$ term to be negative,
confirming our prediction
from holography.

\section{AdS/CFT derivation of the Froissart bound}

In this section, we will review and redo the analysis in~\cite{Giddings:2002cd} relating the gravity calculations
to the Froissart bound in order to find the effects of finite coupling. A more careful analysis of this can be
performed using the results in~\cite{Kang:2005bj}. Since the functional form of the transcendental equation
which gives the maximal impact parameter is similar to what we will use, we do not expect that
the conclusions we arrive at would 
change significantly in this case.

\subsection{Action and equations of motion}

The most general quadratic gravity action in 5 dimensions is given by 
\begin{equation}\label{action1}
S=S_{EH}+S_{GB},
\end{equation}
where $S_{EH}$ is the standard Einstein-Hilbert action with a cosmological constant $\Lambda$
\begin{eqnarray}
S_{EH}=\int{d^5x\sqrt{-\tilde{g}}\left[\frac{1}{\kappa}
\left(\tilde{R}-2\Lambda\right)+\tilde{\mathcal{L}}_{M}\right]}
\end{eqnarray}
and $S_{GB}$ is the Gauss-Bonnet higher derivative correction to it given by 
\begin{eqnarray}
S_{GB}=\frac{\gamma}{\kappa}\int{d^5x\sqrt{-\tilde{g}}\left( \tilde{R}^2-4
\tilde{R}^{ab}\tilde{R}_{ab}+\tilde{R}^{abcd}\tilde{R}_{abcd}\right)},
\end{eqnarray}
with $\kappa\equiv 16\pi G^{(5)}$, $\gamma$ being a perturbative coupling. 
The 5-dimensional gravitational constant $G^{(5)}$ is  related to the Planck mass $M_p$ as
\begin{equation}
G^{(5)}\equiv \frac{1}{M_p^3}.
\end{equation}
$\tilde{g}_{ab}$ is the metric and $\tilde{g}=\det( \tilde{g}_{ab}).$
$\tilde{R}$ is the Ricci scalar,  $\tilde{R}_{ab}$ is the 5-dimensional Ricci tensor and $\tilde{R}_{abcd}$ is
the 5-dimensional Riemann tensor.
The term $\tilde{\mathcal{L}}_{M}$ is the contribution due to any matter fields in the theory.

Extremizing this action gives the 5-dimensional equations of motion
\begin{equation}
\label{EOMweyl}
\tilde{G}_{ab}+\tilde{g}_{ab}\Lambda+\gamma\tilde{B}_{ab}={\kappa}\tilde{T}_{ab},
\end{equation}
where $\tilde{G}_{ab}\equiv \tilde{R}_{ab}-\frac{1}{2}\tilde{R}\tilde{g}_{ab}$ is the Einstein tensor
and $\tilde{B}_{ab}$ is defined as
\begin{equation}
\begin{array}{llll}
\tilde{B}_{ab}&\equiv& -\frac{1}{2}\tilde{g}_{ab}\left(\tilde{R}_{cdef}\tilde{R}^{cdef}
-4\tilde{R}_{cd}\tilde{R}^{cd}+\tilde{R}^2\right)
+2\tilde{R}\tilde{R}_{ab} \\ 
&&+2\tilde{R}_{acde}{\tilde{R}_b}^{\,\,\, cde}
-4\tilde{R}^d_{\,\,\,acb}\tilde{R}_d^{\,\,\,c}-4\tilde{R}_{ac}\tilde{R}_b^{\,\,\,c}.
\end{array}
\end{equation}
The stress-energy tensor $\tilde{T}_{ab}$ is generated by any matter present and  is given by
\begin{equation}
\tilde{T}_{ab}\equiv\frac{1}{\sqrt{-g}}\left(-\frac{\delta \tilde{\mathcal{L}}_M}{\delta \tilde{g}^{ab}}
+\frac{1}{2}\tilde{g}_{ab}\tilde{\mathcal{L}}_M\right).
\end{equation}

\subsection{Solution as background and perturbations}

To solve the equations of motion (\ref{EOMweyl}), we introduce the metric ansatz
\begin{equation}
\tilde{g}_{ab}=g_{ab}+H_{ab},
\end{equation}
 where $H_{ab}$ are the perturbations on an  AdS$_5$ background metric $g_{ab}$ given by
\begin{eqnarray}
g_{ab}=\left(\frac{L}{z}\right)^2
\left(
\begin{array}{c|c}
1 & 0 \\
 \hline 0 & \eta_{\mu\nu}
\end{array}
\right), \quad 
\eta_{\mu\nu}=diag(-1,1,1,1).
\end{eqnarray}
$z$ is the radial coordinate of the AdS$_5$ and $z=L$ is its IR boundary: $0\leq z\leq L$. We use
the convention that indices with greek alphabets
($\mu,\nu,...$) range from 0 to 3 and are raised (lowered) with the metric $\eta^{\mu\nu}$ ($\eta_{\mu\nu}$).
Indices with latin alphabets ($a,b,...$)
 include the radial coordinate $z$ as well and are raised (lowered) with $g^{ab}$ ($g_{ab}$).
The background metric $g_{ab}$ satisfies the Einstein's field equations with a negative cosmological
constant $\Lambda=-\frac{6}{L^2}$:
\begin{eqnarray}
\label{AdSEOM}
R_{ab}-\frac{1}{2}Rg_{ab}-\frac{6}{L^2} g_{ab}=0.
\end{eqnarray}

We redefine the perturbations to a convenient form
\begin{equation}
H_{ab} = \left(\frac{L}{z}\right)^2 h_{ab}
\end{equation}
and assume that the perturbations are small enough to set
\begin{equation}
(h_{ab})^2 \approx 0.
\end{equation}
We further assume that 
\begin{equation}
h_{zz}=0=h_{\mu z}.
\end{equation}
Therefore we are left with ten nonzero components of $h_{ab}$. But there are 15 equations of motion
and hence we have the freedom to choose  the traceless-transverse gauge
\begin{equation}
\nabla^\mu h_{\mu\nu}=0, \quad h\equiv \eta^{\mu\nu} h_{\mu\nu}=0,
\end{equation}
where $\nabla^\mu$ is the covariant derivative in the AdS$_5$ background.

The equations of motion for the perturbation can be derived from  (\ref{EOMweyl}) using (\ref{AdSEOM}). 
In the traceless-transverse gauge,  the linearised equations of motion for $h_{\mu\nu}$ are
\begin{equation}
\label{EOMlinhhat}
\square h_{\mu\nu}=-\frac{\kappa}{C}\left(\frac{z}{L}\right)^2T_{\mu\nu}, \quad \quad C\equiv1-
\frac{4\gamma}{L^2}.
\end{equation}
where $\square\equiv g^{ab}\nabla_a\nabla_b$, using the unperturbed metric.

Thus the effect of the $1/N$ correction will be to simply rescale the Newton constant. 
AdS/CFT arguments suggest that $\gamma>0$ and so, $C<1$~\cite{Buchel:2008vz}, although $\gamma<0$ cannot
be ruled out. Moreover, in this theory, the first $1/\lambda$ correction arises at eight derivative order
through $R^4$ corrections (see eg.~\cite{quantum}), the result of which from the gravity side will 
be further subleading. Since we are treating the curvature corrections perturbatively,
we can in fact give a stronger argument for the generality of our result based on~\cite{ss}.
Notice that in the arguments above, we need to find the linearised equations around AdS.
In~\cite{ss} it was shown that a very general higher curvature action can be rewritten in the form of
a background field expansion around AdS. When this is done, the linearised equations of motion will
only be sensitive to the quadratic curvature terms in the background expanded action.
Since we are working perturbatively, we can use field redefinitions to get back the Gauss-Bonnet gravity.
Hence our claim above will work quite generally with 
the 
coefficient $C$ related to the two point function of the stress tensors in the field theory. In the general
case, $C$ can be either $>1$ or $<1$  so long as it is positive, which is demanded by unitarity of the field
theory.

The rest of the analysis thus parallels~\cite{Giddings:2002cd}, which we will review below in order to
understand the systematics of the subleading terms in the Froissart bound.

\section{Solution to the perturbations}

We are interested in solving for  the linearised perturbations  $h_{\mu\nu}$ which satisfy
(\ref{EOMlinhhat}) with a point mass $m$ on the
IR boundary 
$z=L$. This point mass is a source and its contribution in the action  is accounted for
by adding a suitable  $\mathcal{L}_M$.
The addition of this mass leads to a stress tensor with only one  non-zero component 
\begin{equation}
T_{ab}=\left\{
\begin{array}{lll}
{\textrm{non-zero    }} & \quad\quad\textrm{if }\,\, a=b=0, \\
0 & \quad\quad\textrm{otherwise}.
\end{array}\right.
\end{equation}
But such a $T_{ab}$ is not traceless
\begin{equation}
g^{ab}T_{ab}\neq0
\end{equation}
and hence is not compatible with the  equation of motion  (\ref{EOMlinhhat}).
In order to make it traceless, we also add an incompressible gas  on the brane, which only generates a
pressure $T_{11}$ and no shear stress
($T_{ij}=0$ for $i\neq j$). The resultant traceless stress-energy tensor is
\begin{equation}T_{ab}=\left.
\begin{array}{lll}
m\delta ^3(\overset\rightarrow{x} )\delta\left(z-L\right)\left(\delta_a^0\delta_b^0+\delta_a^1\delta_b^1\right). 
\end{array}\right.
\end{equation}
With this source,
the only non-trivial equations of motion are
\begin{eqnarray}
\square h_{ii}=- \frac{m\kappa}{C} \left(\frac{z}{L}\right)^2
\delta ^3(\overset\rightarrow{x})\delta\left(z-L\right), \quad \quad i=0,1.
\label{ttEOM}
\end{eqnarray}

But before we solve  (\ref{ttEOM}), we need to specify the boundary conditions. As the $h_{\mu\nu}$'s
are gravitational fields,
they must satisfy Neumann boundary conditions in the IR brane
\begin{equation}
\label{NeuBC}
\left.n^I\partial_Ih_{\mu\nu}\right|_{z=L}=0,
\end{equation}
where $n^I$ is the unit vector outward, normal to the IR boundary.

The boundary problem (\ref{ttEOM}) 
can be solved using the scalar Green function $\Delta_{ii}^{(5)}\left(X,X^{\prime}\right)$ satisfying
\begin{eqnarray}
\square\Delta_{ii}^{(5)}\left(X,X^{\prime}\right)=
\frac{1}{\sqrt{-g}}\delta ^4\left(x-x^{\prime} \right)\delta\left(z-z^{\prime}\right),\label{ecuacionprima} 
\end{eqnarray}
with coordinates $X=\left(z,x\right)$. The boundary condition (\ref{NeuBC}) translates to
\begin{eqnarray}
\label{condicionprima}
\left.\partial_z\Delta_{ii}^{(5)}\left(X,X^{\prime}\right)\right|_{z=L}=0.
\end{eqnarray}
 The perturbations $h_{ii}$ can be obtained from $\Delta_{ii}^{(5)}$ as follows:
\begin{eqnarray}\label{h00hatgreen}
h_{ii}= 
-\frac{m\kappa}{C}  \int{d^5X^\prime
\Delta_{ii}^{(5)}\sqrt{-g}\left(\frac{z^{\prime}}{L}\right)^2\delta ^4(\overset
\rightarrow{x}^{\prime},z^\prime-L)},
\end{eqnarray}
where
\begin{equation}
\delta ^4(\overset
\rightarrow{x}^{\prime},z^\prime-L)=\delta ^3(\overset
\rightarrow{x}^{\prime})\delta(z^\prime-L).
\end{equation}

Even though (\ref{ecuacionprima}) is a second order partial differential equation, in the Fourier space
it reduces to a second order ODE in the variable $z$.
The Fourier transform in Minkowski 4 dimensions defined as
\begin{equation}
\label{FTec}
\Delta_{ii}^{(5)}\left(X,X^{\prime}\right)\equiv
\int{\frac{d^4p}{\left(2\pi\right)^4}e^{ip\left(x-x^{\prime}\right)}
\Delta_p\left(z,z^{\prime}\right)}
\end{equation}
 must satisfy
\begin{equation}
\label{miecu}
\frac{1}{L^2}\left(z^2\partial_z^2-3z\partial_z+q^2z^2\right)\Delta_p=
\left(\frac{z}{L}\right)^5\delta\left(z-z^{\prime}\right),
\end{equation}
with $q^2\equiv -p^2$. 
Under the redefinition 
\begin{equation}
\Delta_p = \left(\frac{zz^\prime}{L^2}\right)^2 \hat{\Delta}_p,
\end{equation}
equation (\ref{miecu}) becomes
\begin{equation}
\label{miecu1}
\left(z^2\partial_z^2+z\partial_z+q^2z^2-4\right)\hat{\Delta}_p=L
\frac{z^3}{{z^\prime}^2}\delta\left(z-z^{\prime}\right).
\end{equation}
For $z\neq z^{\prime}$, the above equation admits as its two independent solutions the Bessel
functions $J_2\left(q z\right)$ and
$Y_2\left(qz\right)$:
\begin{eqnarray}
\hat{\Delta}_p\left(z,z^{\prime}\right)=
\begin{cases}
\hat{\Delta}_+\left(z,z^{\prime}\right) \quad \textrm{if }  z>z^{\prime} \\
\hat{\Delta}_-\left(z,z^{\prime}\right) \quad \textrm{if } z<z^{\prime},
\end{cases}
\end{eqnarray}
where
\begin{equation}
\hat{\Delta}_{\pm}\left(z,z^{\prime}\right)=A_\pm\left(z^{\prime}\right)J_{2}\left(qz\right)
+B_\pm\left(z^{\prime}\right)Y_{2}\left(qz\right).
\end{equation}
$A_\pm$ and $B_\pm$ are functions of $z^{\prime}$ and can be determined from the boundary and matching conditions.

In the region $z>z^\prime$, the boundary condition (\ref{condicionprima}) sets
\begin{equation}
A_+=-B_+\frac{Y_1(qL)}{J_1(qL)}.
\end{equation}
In the region $z<z^{\prime}$, demanding that  $\Delta_{ii}^{\left(5\right)}$ be regular at $z=0$ yields 
\begin{equation}
B_-=0.
\end{equation}

Equation (\ref{miecu1}) implies $\hat{\Delta}_p$ is continuous across $z=z^{\prime}$
\begin{equation}
\label{match1}
\left.\hat{\Delta}_+\left(z,z^{\prime}\right)\right|_{z=z^{\prime}}=
\left.\hat{\Delta}_-\left(z,z^{\prime}\right)\right|_{z=z^{\prime}},
\end{equation}
but its derivative in $z$ is not. The value of the discontinuity in
$\partial_z\hat{\Delta}_p$ is easily found by integrating (\ref{miecu1}) across $z=z^{\prime}$:
\begin{eqnarray}
\label{match2}
&&\left. \partial_z\left[\hat{\Delta}_+\left(z,z^{\prime}\right)-\hat{\Delta}_-
\left(z,z^{\prime}\right)\right]\right|_{z=z^{\prime}}=
\frac{L}{z^\prime}.
\end{eqnarray}
The matching conditions (\ref{match1}) and (\ref{match2}) yield 
\begin{eqnarray}
\begin{array}{llll}
&&A_-=\frac{\pi  L}{2}
\frac{J_1(qL)Y_2(qz^\prime)-J_2(qz^\prime)Y_1(qL)}{J_1(qL)},\\ 
&&B_+=\frac{\pi  L}{2}J_2\left(qz^{\prime}\right).
\end{array}
\end{eqnarray}
Consequently, we can express $\Delta_p$ as
\begin{eqnarray}
\nonumber
\Delta_p&=&\frac{\pi  L}{2}\left(\frac{z_<z_>}{L^2}\right)^2 \frac{J_2 (qz_<)}{J_1(qL)}\\
&&\left[J_1(qL)Y_2(qz_>)
-Y_1(qL)J_2 (qz_>)\right],
\end{eqnarray}
where $z_<$ $\left(z_>\right)$ denotes the lesser (greater) of $z$ and $z^{\prime}$. 

The Green function $\Delta_{ii}^{\left(5\right)}\left(X,X^{\prime}\right)$ can be obtained by the inverse
Fourier transform
\begin{eqnarray}
\Delta_{ii}^{\left(5\right)}\left(X,X^{\prime}\right)&= &
\frac{\pi L}{2}\left(\frac{{z_<z_>}}{L^2}\right)^2
\int{\frac{d^4p}{\left(2\pi\right)^4}e^{ip\left(x-x^{\prime}\right)}\frac{J_2 (qz_<)}{J_1(qL)}} \nonumber \\
&&\left[J_1(qL)Y_2(qz_>)
-Y_1(qL)J_2 (qz_>)\right]. \label{greenlin}
\end{eqnarray}
In the case where one of the arguments of $\Delta_{ii}^{\left(5\right)}$ is on the IR brane, at $z^{\prime}=L$,
one can use
the Wronskian of the Bessel functions
\begin{equation}
\label{usewrons}
\begin{array}{lll}
Y_n(qL)J_{n+1}(qL)-J_{n}(qL)Y_{n+1}(qL)=\frac{2}{\pi qL}, 
\end{array}
\end{equation}
valid for all $n \in \mathbb{N}$ to reduce (\ref{greenlin}) to
\begin{eqnarray}
\Delta_{ii}^{\left(5\right)}\left(X,L,x^{\prime}\right)=-\left(\frac{z}{L}\right)^2
\int{\frac{d^4p}{\left(2\pi\right)^4}e^{ip\left(x-x^{\prime}\right)}}\frac{J_{2}\left(qz\right)} {qJ_1(qL)}.
\end{eqnarray}
Using (\ref{h00hatgreen}) and integrating over $x^{\prime}$ and $p^0$ we get
\begin{eqnarray}
\label{expression}
\hspace*{-0.3cm}h_{ii}\left(X\right)=\frac{m\kappa}{C}\left(\frac{z}{L}\right)^2
\int\frac{d^3p}{(2\pi)^3}e^{i\vec{p}\cdot\vec{x}}
\frac{J_{2}\left(|\vec{q}|z\right)} {|\vec{q}|J_1(|\vec{q}|L)}.
\end{eqnarray}

Although the integral in (\ref{expression}) is difficult to evaluate explicitly, we are only interested in
the long-distance $z<<L$ limit where it simplifies.
In this case, the integral is dominated by the region of small $|\vec{q}|z$ and we can replace the Bessel
function
$J_{2}\left(|\vec{q}|z\right)$ by a small argument expansion. We find
\begin{eqnarray}
h_{ii}\left(X\right)\approx
\frac{m\kappa}{8C}\left(\frac{z^2}{L}\right)^2\int{\frac{d^3p}{(2\pi)^3}e^{i\vec{p}\cdot\vec{x}}
\frac{|\vec{q}|}{J_1\left(|\vec{q}|L\right)}}.
\end{eqnarray}
In the spherical coordinates
\begin{equation}
d^3p=2\pi |\vec{p}|^2d|\vec{p}|d\left(\cos\theta\right),
\end{equation}
integration over the angular dependence yields
\begin{eqnarray}
\nonumber
&&h_{ii}\left(X\right)\approx \\ \label{tobeintegrated}
&&\frac{m\kappa}{32\pi^2C}\left(\frac{z^2}{L}\right)^2\frac{1}{r}
\int_0^\infty{d|\vec{p}|\cdot |\vec{p}|^2
\frac{ e^{i|\vec{p}|r }-e^{-i|\vec{p}|r}}{J_1\left(|\vec{q}|L\right)}},
\end{eqnarray}
where $r\equiv |\vec{x}|$. 

The denominator of (\ref{tobeintegrated}) has zeros where $|\vec{q}|L$ satisfies
\begin{equation}
\label{denominador}
J_1\left(|\vec{q}|L\right)=0.
\end{equation}
There are infinite number of such zeros and hence the integrand has infinite number of first order poles.
These poles are located on the positive real axis of the $|\vec{q}|$-plane:
\begin{equation}
|\vec{q}|_{pole}= \frac{j_{1,k}}{L} , \quad j_{1,k}\in\mathbb{R}^+, \quad k=1,2,\ldots
\end{equation}
It is easy to see that in the long-distance (large $r$) limit, the contributions from the higher poles in
the integral
(\ref{tobeintegrated}) are exponentially suppressed and hence can be ignored. Thus, it suffices to consider
the first  pole
at $|\vec{q}|L=j_{1,1}$, 
but we will keep the second pole at $|\vec{q}|L=j_{1,2}$  as well to explicitly check the above claim.

The denominator of (\ref{tobeintegrated}) can be expanded in the small neighbourhood of the poles, where
the integrand is expected to contribute significantly. 
Using the asymptotic behaviour of the Bessel functions for a small argument, we can write 
\begin{equation}
\label{resexp}
\lim_{|\vec{q}| \rightarrow j_{1,k}} J_1\left(|\vec{q}|L\right)\approx \frac{|\vec{q}|L-j_{1,k}}{2}.
\end{equation}
This property can be used to evaluate the residues at the poles $j_{1,1}$ and $j_{1,2}$.
The integral can thus be evaluated to
\begin{eqnarray}
\nonumber
\hspace*{-0.1cm}h_{ii}&\approx&\frac{m\kappa (j_{1,1})^2}{16\pi CL }
\left(\frac{z}{L}\right)^4\frac{e^{-\frac{j_{1,1}r}{L}}}{r}\\
&&\left[1+
\left(\frac{j_{1,2}}{j_{1,1}}\right)^2e^{-\frac{\left(j_{1,2}-j_{1,1}\right)r}{L}}\right].\label{hzerozerozero}
\end{eqnarray}
The second term in the parenthesis is the contribution from the second pole $j_{1,2}$. As previously suggested,
this term is exponentially suppressed as
compared
to the leading term in $j_{1,1}$ and hence can be safely ignored  in the large $r$ limit.

\section{Gauge/gravity duality: upper bound to the scattering cross section}

A sufficiently large mass $m$ 
in (\ref{hzerozerozero}) will make the perturbation big and when $h_{ii}\sim 1$ we can expect a
black hole to be formed~\cite{Giddings:2002cd}. 
The causal horizon of the black hole lies at $r=r_H$, where 
 $H_{00}=1$ (where $\tilde{g}_{00}$ vanishes).

So, in the IR boundary at $z=L$, an estimate for $r_H$ can be obtained from (\ref{hzerozerozero}) as
\begin{eqnarray}
\frac{m\kappa (j_{1,1})^2}{16\pi CL }\frac{e^{-\frac{j_{1,1}r_H}{L}}}{r_H} \propto \rho,\label{totalhis1}
\end{eqnarray}
with $\rho$ a constant of order 1: $\rho\sim\mathcal{O}(1)$.
Identifying the black hole energy with its mass, $E\propto m$,
we can re-express (\ref{totalhis1})  as
\begin{eqnarray}
\label{radius}
r_H\approx \frac{1}{M_1}\left(\ln\frac{E}{E_0}-\ln\frac{r_H}{L}-\ln{\hat{C}}\right),
\end{eqnarray}
where $E_0\equiv \frac{16\pi L^2}{\kappa  (j_{1,1})^2}$ has the dimension of energy,
$M_1\equiv \frac{j_{1,1}}{L}$ is the mass of the
lightest Kaluza-Klein mode of the graviton and $\hat{C}\gtrsim\mathcal{O}(1)$ is a constant related to $C$
which absorbs the ambiguity in the RHS of (\ref{totalhis1}).

To solve the transcendental equation (\ref{radius}), we insert the ansatz
\begin{equation}
\label{ansatz}
r_H=\frac{1}{M_1}\left(\ln\frac{E}{E_0}-\ln{\hat{C}}\right)+c_1\ln\left(c_2\ln\frac{E}{E_0}\right),
\end{equation}
with $c_1,c_2$ constants to be determined. In the long-distance, high-energy ($r,E$ large) limit,
(\ref{ansatz}) satisfies the equation
with $$c_1=-\frac{1}{M_1}, \quad c_2=\frac{1}{M_1 L}.$$
The geometric cross section of the black hole is then given by
\begin{equation}
\sigma_{BH}= \omega \pi r_{H}^2,
\end{equation}
where $\omega$ is some constant different from unity which takes into account
the higher curvature effects through the Wald formula. We can absorb this factor into $1/M_1^2$, which
sits outside our formulae -- this does not play any significant role in our analysis of the general structure.

This gives an estimation of  the maximal scattering cross section in the
gauge theory~\cite{Giddings:2002cd}. The cross section in the gauge theory is
bounded from above by the geometric cross section of
the black hole:
\begin{equation}
\sigma\lesssim\sigma_{BH}.
\end{equation}
Since we want to use the standard notation $s=E^2$ and normalise using $s_0=2m_p^2$, with $m_p$ being
the proton mass, we can set $(E/E_0)^2=(s/s_0)e^{\zeta}$ to get
\begin{equation}\label{longsigma}
\sigma\lesssim \frac{\pi}{M_1^2}\left( \frac{1}{2}\ln \frac{s}{s_0}+\frac{\zeta}{2}-\ln \hat{C}-
\ln(\frac{1}{2M_1 L}[\ln\frac{s}{s_0}+\zeta])\right)^2\,.
\end{equation}
The first zero of (\ref{denominador}) is at $j_{1,1}\approx3.8$,
and so $M_1 L\approx 3.8$.

\section{Corrections to the Froissart bound}

As we pointed out in the introduction, the $(\ln\frac{s}{s_0})$ term has  been widely used
in fits to experimental data. Recent developments in the theory suggest
that a $(\ln\frac{s}{s_0})(\ln\ln\frac{s}{s_0})$-like term might be necessary in addition at higher
energies~\cite{Martin:2013xcr}.

In the high energy (large $s$) limit (\ref{longsigma}) reduces to
\begin{equation}
\label{shortsigma}
\sigma\lesssim\frac{\pi}{M_1^2}\left[\frac{1}{4}\ln^2\frac{s}{s_0}+\beta \ln\frac{s}{s_0}-
(\ln\ln\frac{s}{s_0})\ln\frac{s}{s_0}\right].
\end{equation}
This is our main result, which we will use in the fits to experimental data below.
The sign of the coefficient $\beta$ needs to be determined next.  Here,
\begin{equation}
\beta=\frac{1}{2}\ln\left[8\left(\frac{m_p}{M_p}\right)^2\left(\frac{M_1}{M_p}\right)^4\left(\frac{M_1L}{\hat{C}}
\right)^2\right].
\end{equation}
In gauge/gravity duality, $M_p^3=2N^2/(\pi L^3)$ where $N$ is large and thus $m_p/M_p \ll 1$.
Further $M_1 L\approx 3.8$ with $\hat{C}\gtrsim O(1)$, as mentioned earlier. This means that the argument of
the logarithm  is $\sim m_p^2 L^2/N^4 \hat C^2$.  Hence we expect $\beta<0$, since the argument of
the logarithm is $\ll 1$ for large $N$.  Thus the main conclusions from our analysis are that the structure of
the subleading terms which arise from the Einstein dual are not altered at finite $1/N$, $1/\lambda$ and that
the subleading term $(\ln \frac{s}{s_0} )$ should come with a negative coefficient.
Notice that in order to reach this conclusion we just needed $s_0$ to be less than the scale $E_0$
arising naturally in the AdS/CFT calculation and  the finite coupling effects (in $1/N$ and $1/\lambda$) to
contribute perturbatively (so that $\hat C$ does not become drastically small for instance). We have two
predictions from AdS/CFT.
First that the sign of the $(\ln \frac{s}{s_0} )$ is negative and second that the ratio~\cite{ncomment}
between the coefficients
of the $(\ln^2\frac{s}{s_0})$ term to the $(\ln\frac{s}{s_0})(\ln\ln\frac{s}{s_0})$ term is $-1/4$. 
We will now confront experimental data with these predictions.

\section{Fits to experimental data}

\begin{center}
\begin{table*}[ht]
{\small
\hfill{}
\scalebox{1}{
\begin{tabular}{|l|c|c|c|c|}\hline
 &  $\sigma_{1}^\pm$  &  $\sigma_{2}^\pm$ & $\sigma_{3}^\pm$ & $\sigma_{4}^\pm$  \\ \hline
$a$ (mb) &  $36.980$ & $30.692$ & $37.398$ & $31.786$ \\ 
$b$ (mb) &  $38.3\pm1.6$ & $50.0\pm1.5$ & $36.87\pm0.70$ & $45.38\pm0.20$ \\ 
$c$ (mb) &  $-1.42\pm0.12$ & $0$ & $-1.44\pm0.11$ & $0$\\ 
$d$ (mb) &  $0.043\pm0.042$ & $0.153\pm0.048$ & $0$ & $0$\\ 
$e$ (mb) &  $0.276\pm0.011$ & $0.1687\pm0.0075$ & $0.2844\pm0.0076$ & $0.1922\pm0.0017$\\ 
$f$ (mb) &  $-21.5\pm1.0$ & $-18.7\pm1.0$ & $-20.95\pm0.85$ & $-16.78\pm0.71$\\ 
$\alpha$ &  $0.490\pm0.012$ & $0.525\pm0.014$ & $0.496\pm0.010$ & $0.550\pm0.011$ \\ \hline \hline
$\chi^2$ &  $270.535$ & $276.018$ & $270.606$ & $391.754$ \\
$N_{DF}$ &  $342$ & $343$ & $343$ & $344$\\
$\frac{\chi^2}{N_{DF}}$ &  $0.791$ & $0.805$ & $0.789$ & $1.139$ \\ \hline
\end{tabular}}}
\hfill{}
\caption{Values and errors of the fit parameters in (\ref{fit1})-(\ref{fit4}).
The table includes the corresponding values of $\chi^2$, $N_{DF}$ and
$\frac{\chi^2}{N_{DF}}$. \label{paramtable}}
\end{table*}
\end{center}

The experimentally measured total hadronic cross section  for
$pp$ and $p\bar{p}$ scattering events $\sigma_{tot}^{pp/p\bar{p}}$
shows an initial decrease and a later rise with the centre of mass energy $\sqrt{s}$.
An understanding of this form is beyond the scope of perturbative QCD.
The theory of analyticity of the $S$ matrix demands that the hadronic amplitudes be analytic
functions in the complex angular
momentum $J$. In this plane, the singularities do not depend on the scattering hadrons.
Unitarity of partial waves and boundedness of the absorptive part within the Lehman ellipse
lead to the Froissart (upper) bound on $\sigma_{tot}^{pp/p\bar{p}}$. Also,
Regge trajectory exchanges and a Pomeron
exchange seem to play an important role in the description of $\sigma_{tot}^{pp/p\bar{p}}$.

 The behaviour of the total cross section was first successfully parametrised by Donnachie and Landschoff
 (DL)~\cite{Donnachie:1992ny,Donnachie:2004pi}.
 Using Regge-Pomeron theory, they proposed a sum of two powers to describe the then available data:
 \begin{equation}
 \sigma_{\textrm{tot}}=Xs^\epsilon+Ys^{-\eta},
 \end{equation}
where $X,Y,\epsilon, $ and $\eta$ are parameters and $X$ should be equal for $\sigma_{{tot}}^{pp}$ and
$\sigma_{{tot}}^{p\bar{p}}$. But as new data in a higher energy regime became available, the DL parameterisation
became insufficient. Arguments from crossing symmetry, analyticity and unitarity have since been used to modify
the DL parameterisation and provide a satisfactory description of the data (for example,
see~\cite{Martin,Cudell:2001pn,Block:2006xa}).
An understanding of $pp$ and $p\bar{p}$ scattering is not only intrinsically interesting, but also
leads to important insights into $\gamma p$ and $\gamma\gamma$ ($\gamma$ = photon) scattering
processes (eg.~\cite{Block:1998hu,Godbole:2010ix}).

It is obvious that (\ref{shortsigma}) will be a better fit to experimental data than just the
$( \ln^2 \frac{s}{s_0})$ term,
because of the bigger
parameter space available. But the question one should ask is whether the improvement to the fit
is just because there are more parameters in the
theory or whether the inclusion of the correction terms is really necessary to improve the fit. The answer to
this question can be given by
figuring out whether the terms are statistically  significant in the $F$ test.

\subsection{Data fitting}

A good parametrisation of the data separates the cross section into two parts:
a piece which saturates to a constant value and a piece which accounts for the rise~\cite{Godbole:2004kx}.
The first piece contains
a constant term due to the Pomeron and a Regge term decreasing as $1/\sqrt{s}$.
Two such Regge  terms are required,
representing the $C$-even and $C$-odd exchanges.
It is well-known that $\sigma_{tot}^{p\bar{p}}-\sigma_{tot}^{pp}\propto s^{-1/2}$~\cite{Block:1982bv}.
The second piece comes from
a triple pole at $J=1$, which produces $(\ln^2 \frac{s}{s_0})$ and $(\ln \frac{s}{s_0})$ terms in the total
cross section~\cite{Cudell:2001pn}. At large energies, the Regge terms do not contribute. 
The above mentioned logarithmic piece certainly  dominates asymptotically, but plays an important role at
all energy values. This is due to the fact that the fitted functional form has to describe  the early fall
of the cross section with the energy as well as the common  (for $pp$ and $p\bar p$) constant value from
where the high energy rise begins.
Recently, the determination of the energy scale of the logarithmic terms
has been shown to further contribute a $(\ln\frac{s}{s_0})(\ln\ln\frac{s}{s_0})$ term to the total hadronic
cross section~\cite{Martin:2013xcr}~\footnote{ This term is also relevant in the case of
inelastic cross sections~\cite{Martin:2015kaa}.}.
Consequently, 
we have simultaneously fitted $\sigma_{tot}^{pp/p\bar{p}}$
experimental data points
to the following functional form:
\begin{equation}
\label{fit1}
\sigma_{1}^\pm=a+\frac{b}{\sqrt{y}}+c\ln{y}+d\ln{y}(\ln({\ln{y}}))+e\ln^2{y}\pm f
y^{\alpha-1},
\end{equation}
where the upper (lower) sign refers to $pp$
($p\bar{p}$), $y\equiv\frac{s}{2m_p^2}$ and $\left\{a,b,c,d,e,f,\alpha\right\}$ are
fitting
parameters~\footnote{The parameter $\alpha$ is sometimes set to $\alpha=1/2$ (for example in~\cite{Igi:2005jm}.). 
Here we do not assume $\alpha=1/2$, but rather we show from the fits that $\alpha\approx1/2$
(see TABLE \ref{paramtable}).}.
A single set of these parameters accounts for both $pp$ and $p\bar{p}$ data.

\begin{center}	
\begin{figure*}[t]
{\small
\hfill{}
\includegraphics[scale=0.9]{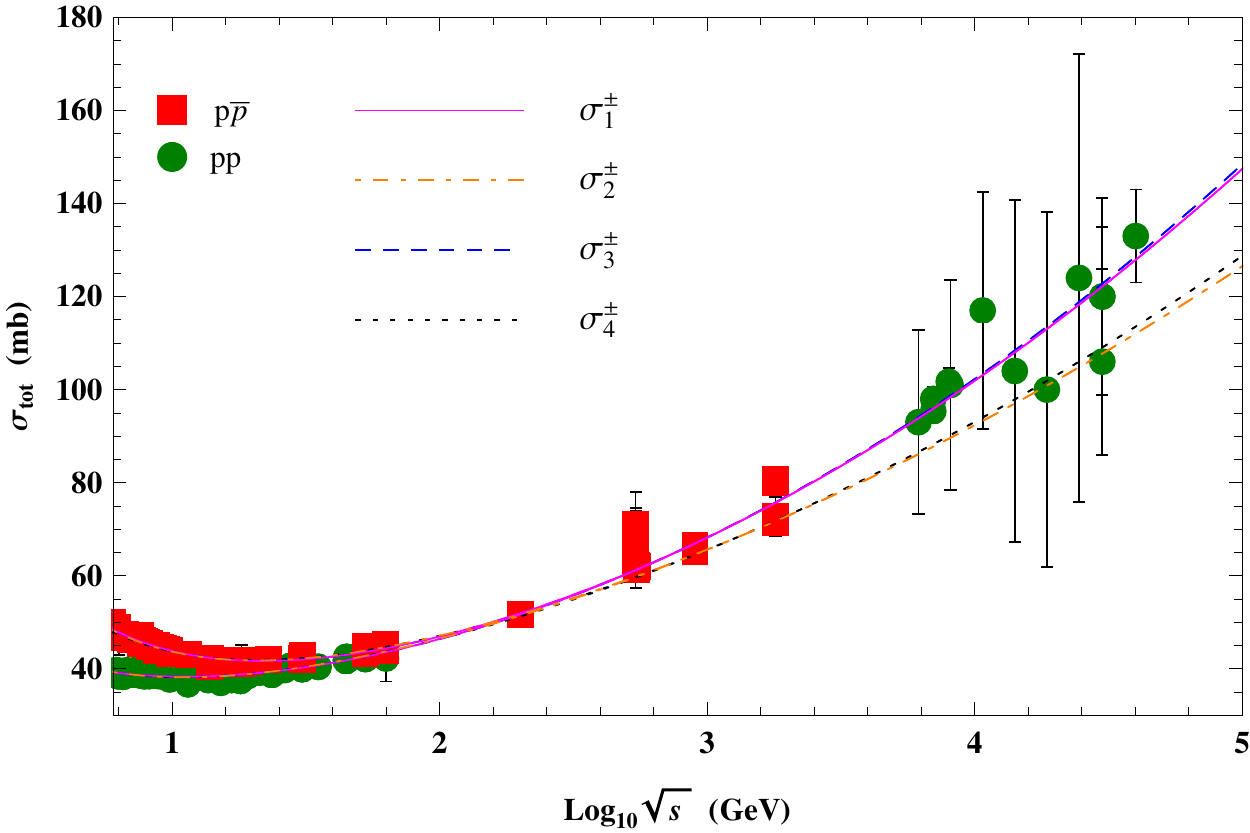}}
\hfill{}
\caption{(Colour online.) Fit results to experimental values of $\sigma_{tot}^{pp}$ and
$\sigma_{tot}^{p\bar{p}}$. The magenta solid, orange dot-dashed, blue dashed and black dotted curves
are the (\ref{fit1})-(\ref{fit4}) fits to the $pp$ (green circles) and $p\bar{p}$ (red squares) data
points, respectively. The data are from CDF, E710, E811, UA1, UA4, UA5
experiments~\cite{Amsler:2008zzb,Battiston:1982su,Amos:1990jh,Abe:1993xx,Abe:1993xy,Avila:1998ej,
Augier:1994jn,Amos:1991bp}.
The $pp$ data points also include   $\sigma_{tot}^{pp}$ results from the LHC
(at $\sqrt{s}=7,8$ TeV)~\cite{TOTEM1,Aad:2014dca,TOTEM2} and cosmic-ray data~\cite{Block:2000pg}.
\label{fitfigure}}
\end{figure*}
\end{center} 

It is important to note that the term $\pm fy^{\alpha-1}$ in (\ref{fit1}) plays a fundamental role in the
simultaneous description of $pp$ and $p\bar{p}$ scattering. This term accounts for the difference
between $pp$ and $p\bar{p}$ scattering in the low energy region while ensuring that a unique set of
parameters is used in the high energy region, where there is little difference between the two sets of
data~\footnote{ Independent fits to $pp$ and $p\bar{p}$ data without $\pm fy^{\alpha-1}$ lead to different values
of the
parameters $c,d,e$ relevant to the description of the data at high energies in the two cases. Since we want
to describe both $pp$ and
$p \bar p$ data with the {\it same} high energy parameters,   we  do not perform such separate fits. It should
be stated however,
that the conclusion about the sign of the $(\ln \frac{s}{s_{0}})$ term that we draw in the end is unchanged
even then.}.

We also consider simpler fits by constraining the value of some of these fitting parameters:
\begin{eqnarray}
\label{fit2}
&&\sigma_{2}^\pm= \sigma_{1}^\pm \quad\textrm{s.t. } c=0,\\
\label{fit3}
&&\sigma_{3}^\pm=\sigma_{1}^\pm \quad\textrm{s.t. } d=0,\\
\label{fit4}
&&\sigma_{4}^\pm=\sigma_{1}^\pm \quad \textrm{s.t. } c,d=0.
\end{eqnarray}
The analyticity of these $\sigma^\pm$'s enforces the constraint~\cite{Block:2005ka}
\begin{eqnarray}
\label{constraintc}
a=48.58-0.3516b-2.091c+2.715d-4.371e.
\end{eqnarray}

The fit (\ref{fit1}) takes the functional form used by~\cite{Block:2005ka} and adds the term
$d\ln{y}(\ln({\ln{y}}))$ to it.
The fits (\ref{fit2})-
(\ref{fit4}) are required to study whether correction terms  found from gauge/gravity duality in the previous
section (see equation (\ref{shortsigma}))
lead to a better description of
the data.

Setting $\sqrt{s}_{\textrm{min}}=6$ GeV allows us to exploit the rich sample of low energy data just above
the resonance
region.

The goodness of the fits has been estimated with the chi-squared value per degree of freedom:
\begin{equation}
\label{chidof}
\frac{\chi^2}{N_{DF}}\equiv\frac{\chi^2}{N_d-N_p},
\end{equation}
where $N_d$ is the number of data points considered and $N_p$ is the number of parameters in the fit.
The chi-squared value
$\chi^2$ has been taken to be
\begin{equation}
\chi^2\equiv\sum_i\left(\frac{\sigma_i^{(\textrm{fit})}-\sigma_i^{(\textrm{exp})}}{\delta\sigma_{i}}\right)^2,
\end{equation}
where $\sigma_i^{(\textrm{fit})}$ is the total cross section given by the fit, $\sigma_i^{(\textrm{exp})}$ is
the experimentally measured total
cross section and $\delta\sigma_{i}$ is the error in the experimental total cross section. 
If the chi-squared value per degree of freedom is of order 1
\begin{equation}
\label{order1}
\frac{\chi^2}{N_{DF}}\sim\mathcal{O}(1),
\end{equation}
then we consider the fit to be good.

The results of the fits ({\ref{fit1})-(\ref{fit4}}) constrained by (\ref{constraintc}) are tabulated in
TABLE \ref{paramtable} . All fits are good, since they satisfy (\ref{order1}).  Note the small errors in
the fit values of $c$. Thus $c$ is indeed negative to a very high degree of significance.

The $pp$ and $p\bar{p}$ cross sections derived from the parameters of TABLE \ref{paramtable} are shown
in FIG. \ref{fitfigure} as a function of the centre 
of mass energy $\sqrt{s}$. The $p\bar{p}$  data points (red squares) include $\sigma_{tot}^{p\bar{p}}$ results
from the experiments CDF, E710, E811, UA1, UA4 and
UA5~\cite{Amsler:2008zzb,Battiston:1982su,Amos:1990jh,Abe:1993xx,Abe:1993xy,Avila:1998ej,
Augier:1994jn,Amos:1991bp}.
The $pp$ data points (green circles) include   $\sigma_{tot}^{pp}$ results from LHC
(at $\sqrt{s}=7,8$ TeV)~\cite{TOTEM1,Aad:2014dca,TOTEM2} and cosmic-ray data~\cite{Block:2000pg}.
The magenta solid, orange dot-dashed, blue dashed and black dotted curves
are the $\sigma_1^{\pm},\sigma_2^{\pm},\sigma_3^{\pm},\sigma_4^{\pm}$ fits to
the $pp$ and $p\bar{p}$ data, respectively.

\subsection{Comparison of the fits}

In this subsection, we study whether and which one of the fits (\ref{fit1})-(\ref{fit4}) provides
a statistically significantly better
description of the experimental data.

The fit (\ref{fit4}) is the simplest of all: it can be obtained from (\ref{fit1})-(\ref{fit3}) by setting
$c=0$ or $d=0$ or both.
Hence, it can be thought of as a special case of any of (\ref{fit1})-(\ref{fit3}). Similarly, (\ref{fit2})
and (\ref{fit3}) can be obtained from
(\ref{fit1}) by setting $c=0,d=0$, respectively.

For this reason, we can consider pairs of such fitting models as nested models. One can compare a pair
of models which are nested by taking
one of them to be the null hypothesis (the model with fewer parameters) and the other one as the alternative
hypothesis (the model with more parameters).
It is obvious that the alternative hypothesis will almost always have a smaller $\frac{\chi^2}{N_{DF}}$, due
to the bigger parameter space available.
In order to determine if such an improvement
in $\frac{\chi^2}{N_{DF}}$ is significant, the $F$ test needs to be used (for a review of the $F$ test,
see~\cite{regressionbook}).

In a $F$ test, the $\chi^2$ and $N_{DF}$ values of the null and alternative hypotheses are used to
compute the $F$ ratio as
\begin{equation}
F\equiv\frac{\left(\chi^{2(\textrm{null})}-
\chi^{2(\textrm{alt})}\right)/\chi^{2(\textrm{alt})}}{\left(N_{DF}^{(\textrm{null})}-
N_{DF}^{(\textrm{alt})}\right)/N_{DF}^{(\textrm{alt})}}.
\end{equation}
If $F\approx1$, then the improvement in $\frac{\chi^2}{N_{DF}}$ provided by the alternative hypothesis
should not be regarded as significant. Rather,
the null hypothesis should be considered sufficient to account for the data. The converse is true if $F>1$.

Since the $F$ distribution is known, using the $F$ ratio and the quantities
\begin{equation}
N_N\equiv N_{DF}^{(\textrm{null})}-N_{DF}^{(\textrm{alt})}, \quad
N_D\equiv N_{DF}^{(\textrm{alt})},
\end{equation}
a $P$ value can be obtained. Such $P$ value quantifies the probability with which one expects a particular
$F$ ratio value (or higher), provided
the null hypothesis is correct. 
If $P<\beta$, with $\beta$ the significance level chosen, then the alternative hypothesis should be
considered to fit the data
significantly better than the null hypothesis, and vice-versa.

TABLE \ref{ftesttable} contains the relevant quantities in the computation of $P$ for the comparison of
all pairs of nested models.
It is easy to see that $\sigma_4^\pm$ must be rejected in favour of any other model, with a significance level
$\beta=0.0001$ when compared to $\sigma_1^\pm$ and $\sigma_3^\pm$ and with a significance level
$\beta=0.002$ when compared to $\sigma_2^\pm$.
$\sigma_1^\pm$ is better than $\sigma_2^\pm$ with a significance level $\beta=0.0001$. There is no
statistically significant difference between $\sigma_1^\pm$ and
$\sigma_3^\pm$.

\begin{table}[H]
\begin{center}
\scalebox{1}{
\begin{tabular}{|c|c|c|c|c|c|} \hline
Null & Alt & $F$ & $N_N$ & $N_D$ & $P$  \\ \hline
$\sigma_2^\pm$ & $\sigma_1^\pm$ & $140.7780$ & $1$ & $342$ & $<0.0001$  \\
$\sigma_3^\pm$ & $\sigma_1^\pm$ & $1.0522$  & $1$ & $342$ & $0.3057$ \\
$\sigma_4^\pm$ & $\sigma_1^\pm$ & $77.3169$ & $2$ & $342$ & $<0.0001$ \\
$\sigma_4^\pm$ & $\sigma_2^\pm$ & $9.8444$ & $1$ & $343$ & $0.0019$ \\
$\sigma_4^\pm$ & $\sigma_3^\pm$ & $153.5580$ & $1$ & $343$ & $<0.0001$ \\ \hline
\end{tabular}
}
\end{center}
\caption{$F$ and $P$ values for the comparison of all possible pairs of nested models. The null hypothesis is
the simpler model. The
alternative hypothesis is the model
containing more parameters.\label{ftesttable}}
\end{table}

 Hence, using the logic of the $F$ test mentioned above, we conclude that 
(\ref{fit3}) provides the best description to the experimental data among the functional forms considered here.
A functional form including a logarithmic term $( \ln \frac{s}{s_0})$ produces a statistically
significantly better description
of the data than just a Froissart-like
bound term alone. The coefficient of this subleading term
is negative. The inclusion of a further subleading term of the form
$(\ln\frac{s}{s_0})(\ln\ln\frac{s}{s_0})$ does not significantly
improve the fit. Hence, as we previously pointed out, the inclusion of more parameters does not necessarily
improve the data description significantly, as it is here the case.

It can be argued that the  $(\ln\frac{s}{s_0})(\ln\ln\frac{s}{s_0})$
correction term may become distinguishable  at higher energies, where it
contributes most. Indeed, in the energy range where data are available, this term behaves almost as a constant
and is subleading to the
$(\ln \frac{s}{s_0} )$ term for the values of $c,d$ found in $\sigma_1^\pm$ (see TABLE \ref{paramtable}). 
More and better data from air shower experiments may be the only possibility of ever addressing the issue
from the point of view of achieving any further discrimination.

\section{Discussions}

The high energy behaviour of the total hadronic cross sections
was first analysed in~\cite{Heisenberg}, where the bound on the rise of the cross section was shown to arise
from the finite range of the interaction.  In the case of QCD, this short range arises from the properties of
the theory in the infrared. This indicates that it is the behaviour of the theory of strong interactions in
the IR that decides the high energy behaviour of the cross section. Arguments from
analyticity and unitarity lead to the Froissart (upper) bound on such total cross sections.
In a Bloch-Nordsiek improved eikonalised mini jet model~\cite{Godbole:2004kx}, the Froissart bound
seems to be directly related to the behaviour of QCD in the far IR.
Phenomenological
fits usually include a subleading logarithmic term as well. A logarithmic subleading term is often used
to fit data, but lacks a theoretical explanation.
It is  just a phenomenological observation.

Ideally, one would like to understand such a subleading term in the context of QCD.
The understanding of the behaviour of total hadronic cross sections  lies in the realm of non-perturbative QCD.
Although there has been a lot of interest in the subject and many efforts have been made, the tools
required for handling  non-perturbative QCD are still being developed. However, using the AdS/CFT
correspondence, one can map a problem in a strongly coupled gauge theory to a problem in a weakly
coupled gravity theory. Hence, we have mapped the high energy behaviour of the total hadronic
cross section to a gravity toy model and investigated what the correction terms to the Froissart
bound are in this simplified scenario.

We have shown that the extension of the holographic arguments in~\cite{Giddings:2002cd,Kang:2004jd} generates 
a subleading $(\ln \frac{s}{s_0} )$  term to the Froissart bound in the dual field theory. 
The duality predicts that the coefficient of such a subleading term is  negative
if the finite coupling corrections are small. In other words, the prediction from
an Einstein dual together with a perturbative $1/N, 1/\lambda$ correction through 
higher curvature corrections is that the coefficient of $(\ln\frac{s}{s_0})$ is negative.
Therefore (since it reduces the upper bound), it is an improvement to the Froissart bound.

Further, we have used all available data to demonstrate that such a $(\ln \frac{s}{s_0})$ term
indeed significantly improves the fits. Also,  the prediction about the negative coefficient for the
subleading term is confirmed from the fits.

 Presumably, the  $(\ln \frac{s}{s_0})$ term reflects the nature of non-perturbative QCD.
While one of course cannot claim that the holographic theory is modelling QCD, it is quite remarkable that
the general structure arising from holography for the subleading terms in the Froissart bound as well as the sign
of the first subleading term agree so well with data.
Hence, it is possible that the gravity toy model here considered captures certain important aspects of
non-perturbative QCD.

While the argument from AdS/CFT leading to the negative coefficient for
the $(\ln \frac{s}{s_0})$ term seems reasonable, another prediction from our AdS/CFT analysis
is that the relative coefficient between the $(\ln^2 \frac{s}{s_0})$ term and the
$(\ln\frac{s}{s_0})(\ln\ln\frac{s}{s_0})$ term is -1/4.
It would be interesting if this was borne
out by the data, but it seems unlikely to happen in the near future.

\mbox{}\\
{\bf Acknowledgements}: V.E.D. would like to thank the Royal
Institute of Technology (Stockholm) and especially Edwin Langmann
for support in an exchange to the Indian Institute of Science (IISc),
where most of the work is done. In IISc, V.E.D. thanks the Centre for High Energy Physics for hospitality.
V.E.D. is also grateful to N.V. Joshi for illuminating discussions and to NSERC and Keshav Dasgupta for financial
support. R.M.G. wishes to acknowledge support from the Department of Science and Technology, India under Grant No.
SR/S2/JCB-64/2007. A.S. acknowledges support from a Ramanujan fellowship, Government of India.

\end{document}